\documentclass[preprint, twocolumn]{aastex631}

\begin{document}

\title{Turbulence properties and kinetic signatures of electron in Kelvin-Helmholtz waves during a geomagnetic storm. }

\author[0000-0001-8698-3383]{Harsha Gurram}
\affiliation{Department of Astronomy, University of Maryland College Park, College Park, Maryland, USA. \\}
\affiliation{NASA Goddard Space Flight Center, Greenbelt, Maryland, USA.}

\author{Jason R. Shuster}
\affiliation{Space Science Center, University of New Hampshire, Durham, New Hampshire, USA.\\}

\author{Li-Jen Chen}
\affiliation{NASA Goddard Space Flight Center, Greenbelt, Maryland, USA.}

\author[0000-0001-6315-1613]{Matthew R. Argall}
\affiliation{Space Science Center, University of New Hampshire, Durham, New Hampshire, USA.\\}

\author{Richard E. Denton}
\affiliation{Department of Physics and Astronomy, Dartmouth College, Hanover, New Hampshire, USA.}

\author{Rachel C. Rice}
\affiliation{Department of Astronomy, University of Maryland College Park, College Park, Maryland, USA. \\}
\affiliation{NASA Goddard Space Flight Center, Greenbelt, Maryland, USA.}

\author{Brandon L. Burkholder}
\affiliation{NASA Goddard Space Flight Center, Greenbelt, Maryland, USA.}
\affiliation{University of Maryland Baltimore, Baltimore, Maryland, USA. \\}

\author{Daniel J. Gershman}
\affiliation{NASA Goddard Space Flight Center, Greenbelt, Maryland, USA.}

\newcommand{\scr}[1]{_{\mbox{\protect\scriptsize #1}} }

\begin{abstract}
The Kelvin-Helmholtz  instability (KHI), in its non-linear phase, plays a significant role in transporting solar wind plasma into Earth’s magnetosphere. This paper presents a comprehensive study of Magnetospheric Multiscale (MMS) spacecraft encounter with KHI during a geomagnetic storm, focusing on elucidating key turbulence properties and reconnection signatures observed at the edges of KH vortices. The spectral slope for electric field stays approximately constant for frequencies below the ion cyclotron frequency ($f < f_{ci}$) and exhibits a break around the lower hybrid frequency ($f_{lh}$), indicating wave activity. Furthermore, MMS observes a current sheet accompanied by intense electron jets and features consistent with strong guide-field asymmetric reconnection across the magnetopause. Substantial agyrotropy (by a factor of $10$) in electron distribution functions is observed in the reconnecting current sheet and at the edges of KH. Our observation presents a multi-scale view into KH turbulence under strongly driven conditions and into the dynamics occurring at electron dissipation scales.  
\end{abstract}

\section{Introduction} \label{sec:intro}
The Kelvin-Helmholtz instability (KHI) is a classical instability that arises at the interface between two fluids with non-zero relative velocity, and occurs in various plasma environments, including at the edge of coronal mass ejections in the solar corona (\cite{foullon_magnetic_2011,Nykyri_Foullon_2013}) and in planetary magnetospheres (\cite{Kivelson1995,Fairfield2003,Delamare2013,hasegawa_transport_2004}). At Earth's magnetopause, this instability can be triggered in the low-latitude magnetopause boundary layer (LLBL) by the significant velocity shear (\cite{LLBL,hasegawa_transport_2004, Taylor, Haaland}). The KH instability on the surface of the magnetopause begins as an anti-sunward surface waves, which 'rolls up' into a train of vortices in the waves' non-linear phase (\cite{nakamura_2021,kavosi_ubiquity_2015}).
These vortices propagate towards the flanks of the magnetosphere and extend into the tail region, playing a crucial role in facilitating the efficient entry of collisionless solar wind plasma into the magnetosphere. KH waves disrupt the frozen-in condition inherent in ideal magnetohydrodynamics (MHD) and play a crucial role in transporting mass and momentum into the magnetosphere through the kinetic process of magnetic reconnection within these vortices (\cite{Nykyri_Otto,nykyri:hal-00317273,li_kinetic_2016,Ma2017,Nykyri2017}).

The first in-situ evidence of rolled-up KH vortices at the Earth's flank magnetopause was reported by Cluster spacecraft, confirming the presence of KH instability under northward solar wind conditions (\cite{hasegawa_kelvin-helmholtz_2009}). These observations showed both cold solar wind particles ($<2$ keV) and hot magnetospheric ions ($>5$ keV) within the vortices, suggest that KHI is the primary mechanism for plasma transport across this boundary. The spacecraft also encountered a current sheet of thickness a few times the ion-inertial length at the trailing edge of the KH wave accompanied with strong Alfv\'{e}nic outflow jets parallel to the magnetosheath flow (\cite{hasegawa_kelvin-helmholtz_2009}). While the Cluster mission could resolve the vortices at ion scales, the Magnetospheric Multiscale (MMS) mission, equipped with high-resolution instrumentation, has offered an unprecedented opportunity to study KH waves at electron scales, enabling examination of small-scale structures and dynamics within the vortices, as well as investigation of micro-physical processes responsible for particle energization and transport. MMS has provided direct kinetic evidence of magnetic reconnection(\cite{eriksson_magnetospheric_2016,eriksson_magnetospheric_2016-1,li_kinetic_2016}) including electron-only reconnection without ion coupling (\cite{phan_electron_2018}) in KH waves (\cite{blasl_multi-scale_2022}). 
Other KH-induced mechanisms observed with MMS are mid-latitude reconnection (\cite{Vernisse2016}) and turbulence (\cite{Starwarz}).

The KHI can drive turbulence in the magnetosheath region, acting as a large-scale mechanism which initiates or reinforces a non-linear cascade (\cite{Turbulence_Hasegawa}). 
 Turbulence within KHI can induce reconnection, causing magnetosheath electrons to traverse the magnetopause and magnetospheric electrons to escape into the magnetosheath  (\cite{eriksson_magnetospheric_2016,li_kinetic_2016}). Observations suggest the presence of Kelvin–Helmholtz waves on both the dawn and dusk flank of the terrestrial magnetosphere (\cite{Otto,Fairfield2003,hasegawa_transport_2004,Starwarz}. The first observations from the MMS of the KHI at the dusk-flank magnetopause during southward interplanetary magnetic field were reported by \cite{blasl_multi-scale_2022}. The measurements show kinetic-scale electric field fluctuations on the low-density side of the edges of waves due to lower-hybrid drift instability (LHDI) excited by the density gradient at the edges of the waves. This instability and fast reconnection disrupt the flow of non-linear KH vortex, leading to a quick decay of the vortex structure possibly explaining the difference in the observation probability of KH waves between northward and southward IMF conditions (\cite{Nakamura2020,Nakamura2022,blasl_multi-scale_2022}).

In this study we analyze KHI event elucidating the key turbulence properties in KH waves during a storm by examining the wave power spectrum. Furthermore, we will discuss the kinetic signatures of electrons near reconnecting current sheets, present at the edges of the KH vortices. The article is organized as follows: In Section~\ref{sec:overview}, we provide an overview of the event under study. In Section~\ref{sec:turbulence}, we analyze magnetic and electric field spectra to understand turbulence properties and energy transfer in KH. In Section~\ref{sec:reconnection}, we present observed signatures of reconnection, and report the key signature of electron agyrotropy observed near reconnecting current sheet. Section~\ref{sec:Agyrotropy} is dedicated to quantifying this agyrotropy and offering insights into its origin. Finally, we draw our conclusions in Section~\ref{sec:Final}.

\section{Overview of MMS observations on 14 April 2022} \label{sec:overview}
\begin{figure*}
\centering
\includegraphics[width=\linewidth]{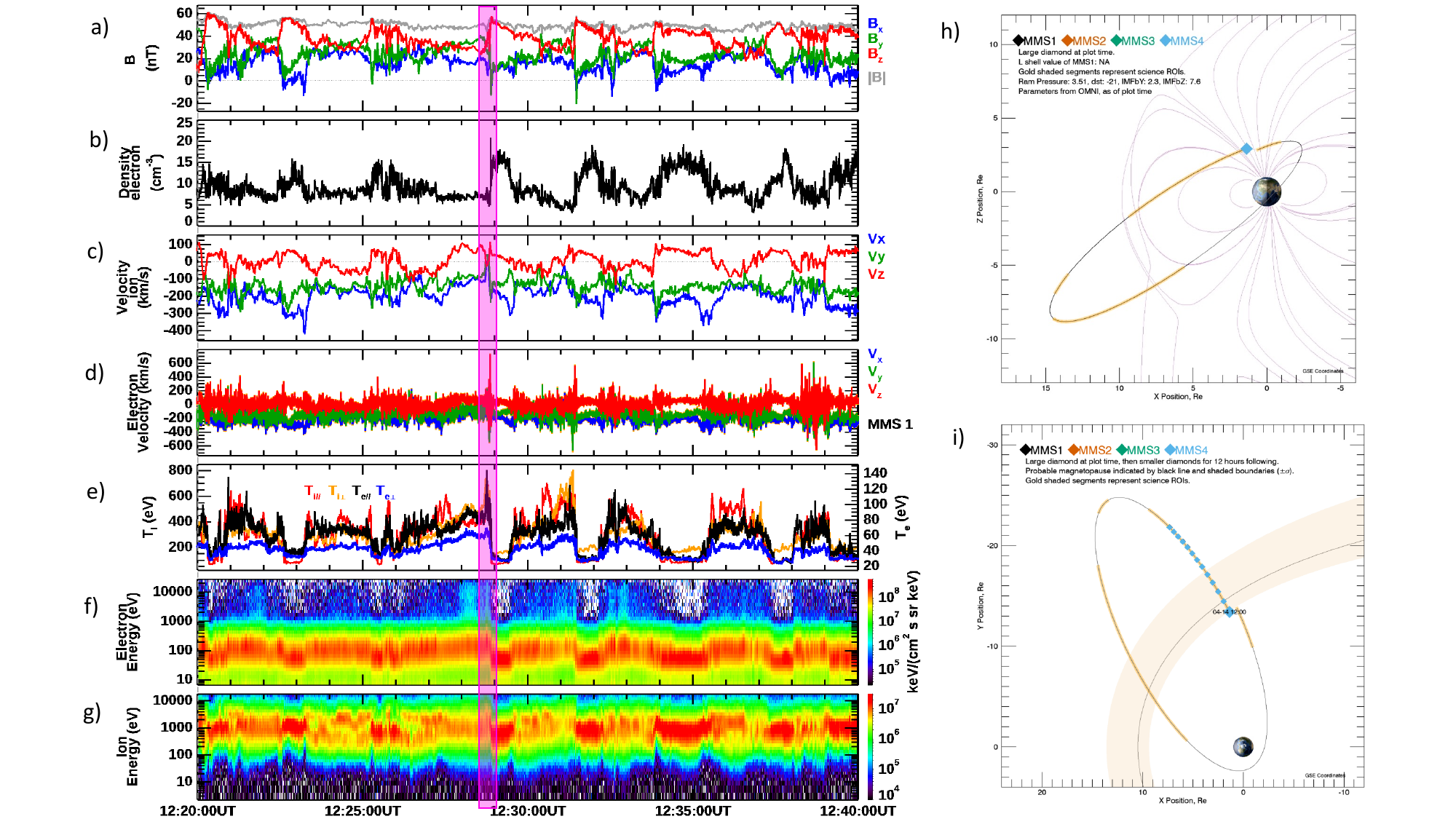}
\caption{Overview of Kelvin-Helmholtz vortices encountered by MMS1 during a geomagnetic storm on 14 April 2022. The panels show a) magnetic field components, b) electron density, c) ion velocity components, d) electron velocity components, e) ion and electron temperatures, f) omnidirectional electron energy spectrum and g) omnidirectional ion energy spectrum. The magenta box represents the reconnection current sheet discussed in Section \ref{sec:reconnection}. An orbital plot of the MMS spacecraft in h) XZ and i) XY plane. All vectors are in GSE coordinates.}
\label{Fig1:overview_plot}
\end{figure*}

During a geomagnetic storm on 14$^{th}$ April 2022, OMNI recorded a decrease in sym-H (minimum of $-86$nT). MMS captured the IMF turning and magnetospheric response from the equatorial dawn flank. MMS made a crossing into the day-side magnetopause at around 10:15UT. From 11:32UT to 12:40UT MMS observes quasi‐periodic high energy populations correspond with fluctuations in density (\cite{Rice2024}). These quasi-periodic fluctuations have been identified as KH waves at the dawn-flank of the magnetopause, exhibiting the characteristics listed in \citet{rice_characteristics_2022}. 
Fig.~\ref{Fig1:overview_plot} displays an overview of the KH interval of interest observed by MMS1. During this period the IMF pointed in the northward direction. Fig.~\ref{Fig1:overview_plot}a) exhibits quasi-periodic fluctuations in the magnetic field component, which are correlated with the fluctuations in electron density and electron temperature shown in Fig.~\ref{Fig1:overview_plot}b) and e) respectively. 
The periodic observation of the magnetosheath and magnetospheric regions is clearly evident in the density and temperatures as MMS alternately encounters regions of plasma from the cold, dense magnetosheath and the hot, tenuous magnetosphere. 

To identify whether the observed fluctuations correspond to KH waves we analyzed the data(not shown) in a LMN system obtained by applying minimum variance analysis on electric field (MVA-E) (\cite{rice_characteristics_2022}). The MVA-E was performed on MMS1 data from 10:45 to 12:45 yielding a global coordinate system given by: $\mathbf{L}_{\scr{GSM}}=[0.5398, 0.5815, 0.6087]$, $\mathbf{M}_{\scr{GSM}}=[0.6048, 0.2351, -0.7609]$ and $\mathbf{N}_{\scr{GSM}}=[0.5856, -0.7789, 0.2247]$.  
Key features aiding in identifying KH waves include semi-periodic fluctuations in magnetosphere and magnetosheath-like plasma properties, such as densities, temperatures, bipolar fluctuations in the normal component of the magnetic field ($B_N$) at the magnetopause, and significant changes in tailward velocity. These fluctuations, typically lasted 3-7 minutes and accompanied by velocity shears, distinguish KH waves from surface waves, with the observed sawtooth pattern and bipolar fluctuations in the $B_N$ component serving as characteristic indicators of KHIs. To differentiate from flux transfer events (FTEs), alignment of pressure minima with $B_N$ is checked, where in KHIs, pressure is lowest in the vortex center where $B_N$ is near $0$, whereas in FTEs, pressure increases when $B_N$ approaches $0$, providing verifiable indicators for KHIs.\\

Fig.~\ref{Fig1:overview_plot} shows MMS1 observations between 12:20:00 UT and 12:40:00UT when the spacecraft encounters the upper boundary layer (\cite{Rice2024}). The average MMS1 position is at $[1.6,-13.9,2.8]R\scr{E}$ in the GSE system. Panels ~\ref{Fig1:overview_plot}h) and ~\ref{Fig1:overview_plot}i) display quasi-periodic perturbations in the omnidirectional energy spectrum of ions and electrons indicating alternating regions of plasma with energies typical of the magnetosheath and magnetosphere, as well as mixed energies due to plasma mixing in the KH vortex. The sharp increase in the magnetic field component $B_z$ from $20nT$ to $60nT$ corresponds to the edges of rolled up KH vortices or the sunward facing edges (e.g., \cite{Otto, li_kinetic_2016, kieokaew_magnetic_2020}). These edges mark the transition between hot and tenuous magnetospheric plasma to cold and dense magnetosheath-like plasma. Some of these crossings exhibit a sharp magnetic field reversals (in $B_y$) confirming the presence of current sheets at the edge of the vortices. One such current sheet is highlighted within a magenta box in Fig.~\ref{Fig1:overview_plot}. It is noteworthy that this specific region also exhibits reconnection signatures, which we will delve into further in Section ~\ref{sec:reconnection}.

\section{Power Spectral Analysis of KHI} \label{sec:turbulence}
\begin{figure*}
\centering
\includegraphics[width=1\linewidth]{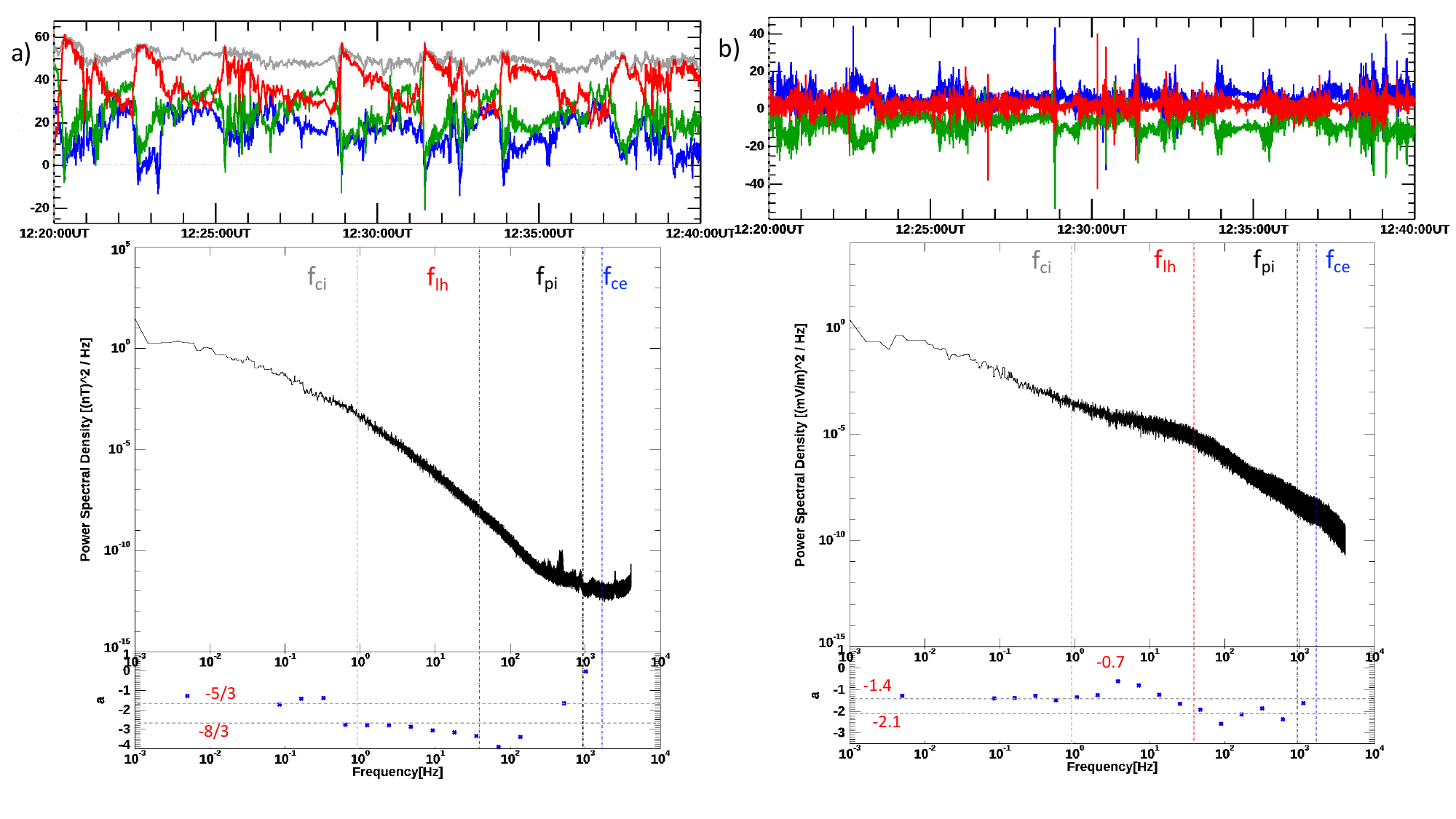}
\caption{ Fluctuations, Power spectra density (PSD) and local slope $a$ for a) magnetic field, $\mathbf{B}$ and b) electric field, $\mathbf{E}$ for the full KHI interval. Vertical dotted lines in grey, red, black and blue indicate the frequencies $f_{ci}$, $f_{lh}$, $f_{pi}$ and $f_{ce}$ respectively. }
\label{Fig4:KH_spectra} 
\end{figure*}
During the investigated interval, the KHI is in its non-linear phase resulting in the breakdown of vortices into smaller structures and initiating energy cascading. Strong turbulence in the KHI is indicated by fluctuations in the magnetic field with $|\delta B_z|/|B| \sim 0.8$. To analyze turbulence properties, we examine the power spectrum of magnetic and electric fluctuations. For the $B$ spectra, we use measurements from Fluxgate-Searchcoil Merged (FSM) (\cite{argall2018fluxgatesearchcoil}) data product created by combining data from the Fluxgate (\cite{Russell2016}) and Search Coil (\cite{LeContel2016}) Magnetometers. For the $E$ spectra the measurements from the instrument Electric field Double probes (EDP) (\cite{Ergun2016a,Lindqvist2016}) is analyzed. Both instruments sample at a frequency of $8192$ samples/second. The characteristic frequencies for this event are ion cyclotron $f_{ci}\approx 1$Hz, lower hybrid $f_{LH}\approx 310$Hz, ion plasma $f_{pi}\approx 1000$Hz  and electron cyclotron $f_{ce}\approx 1500$Hz.\\

Fig.~\ref{Fig4:KH_spectra}(a) and (b) shows the omni-directional magnetic and electric field components and spectra for the KHI interval respectively. The local spectral slope for the spectra was computed from power-law fit in the log–log plane. The trace magnetic fluctuation spectrum follows a power law with the spectral index of $-1.67$ for $f<f_{ci}$ before exhibiting steepening for $f>f_{ci}$. This slope is close to Kolmogorov scaling with $-5/3$ slope.  At the electron scales the local spectra slope has a value $-2.69$ (close to $-8/3$) up to $f_{lh}$. The trace electric field spectrum also follows a power-law and has a spectral index of $-1.4$. Unlike the $B$ spectra, the $E$ spectral slope flattens first at the electron scales and then exhibits steepening. The flattening in the slope occurs between $f_{ci}$ and $f_{lh}$ around $8$Hz with a slope of $-0.7$ and steepens beyond $f_{lh}$ with slope $-2.15$. For higher frequencies the local slope increases due to flattening of the spectrum which is not physical and thought to be due to aliasing of spin tone harmonics (\cite{2013Koval}).

Both $E$ and $B$ spectra obey the theoretical fluid predictions and observations of turbulence in magnetosheath in the low-frequency range $f<f_{ci}$ (\cite{Starwarz,Chen}). The break in the magnetic spectrum happens at $f_{ci}$ which is associated to length scales of $d_i$. This can be related to the scale of reconnecting current sheets in KH, which generally have a thickness $d_i$ (\cite{Leamon_2000,Dmitruk_2004}).  However, at these frequencies, the electric field spectra show a constant spectral slope, suggesting that reconnection in KH vortex has reached a steady state where the electric field is constant while the magnetic field is still reconnecting. The electric field spectrum, however, exhibits a break at frequencies closer to $f_{LH}$. This break indicates a flattening of the spectrum, implying electric energy is not being released like the magnetic energy leading to
a buildup of electric fields and possible gain in electric energy. The significant buildup in electric fields is dissipated by lower hybrid waves possibly, as evidenced by the steepening of the electric spectra beyond $f>f_{lh}$. During magnetic reconnection in a region of strong turbulence in the Earth's magnetotail (\cite{ergun2020,ergun2022}) as well as during KHI with reconnecting current sheets (\cite{Starwarz}) the $B$ spectra slope was $-3.0$ for $f>f_{ci}$ and $E$ spectra slopes were $-0.8$ for $f_{ci} <f <f_{lh}$ and $-2.7\pm 0.2$ for $f>f_{lh}$.  The storm event under discussion exhibits shallower spectral slopes at $f>f_{ci}$, indicating slower dissipation possibly due to the buildup of more electromagnetic energy and fluctuations in KHI. Further investigation will be conducted to understand the effects of geomagnetic storms on the nature of KH turbulence during storms and their impact on the structure and evolution of KH turbulence and its effectiveness in mixing and transport.\\

\section{Reconnection signatures in the Kelvin-Helmholtz vortices}\label{sec:reconnection}
 We investigated the full interval of 20min  for evidence of current sheets with reconnection signatures i.e. ion and electron jets along with field reversals. The 20min burst data showed multiple current sheet crossings at the trailing edge \textit{i.e.} magnetosphere-magnetosheath edge of the KH vortices. This section will focus on the current sheet observed at 12:28:48, highlighted with the magenta box in Fig.~\ref{Fig1:overview_plot}, as it was accompanied by strong electron flow reversals across the current sheet. This particular current sheet stands out in the context of the overall event due to the pronounced anisotropy in temperature observed within the inflow region. Additionally, our analysis reveals the presence of a notable agyrotropy characterized by stretching and rotation in the plane perpendicular to the magnetic field. 

\begin{figure*}
\centering
\includegraphics[width=\linewidth]{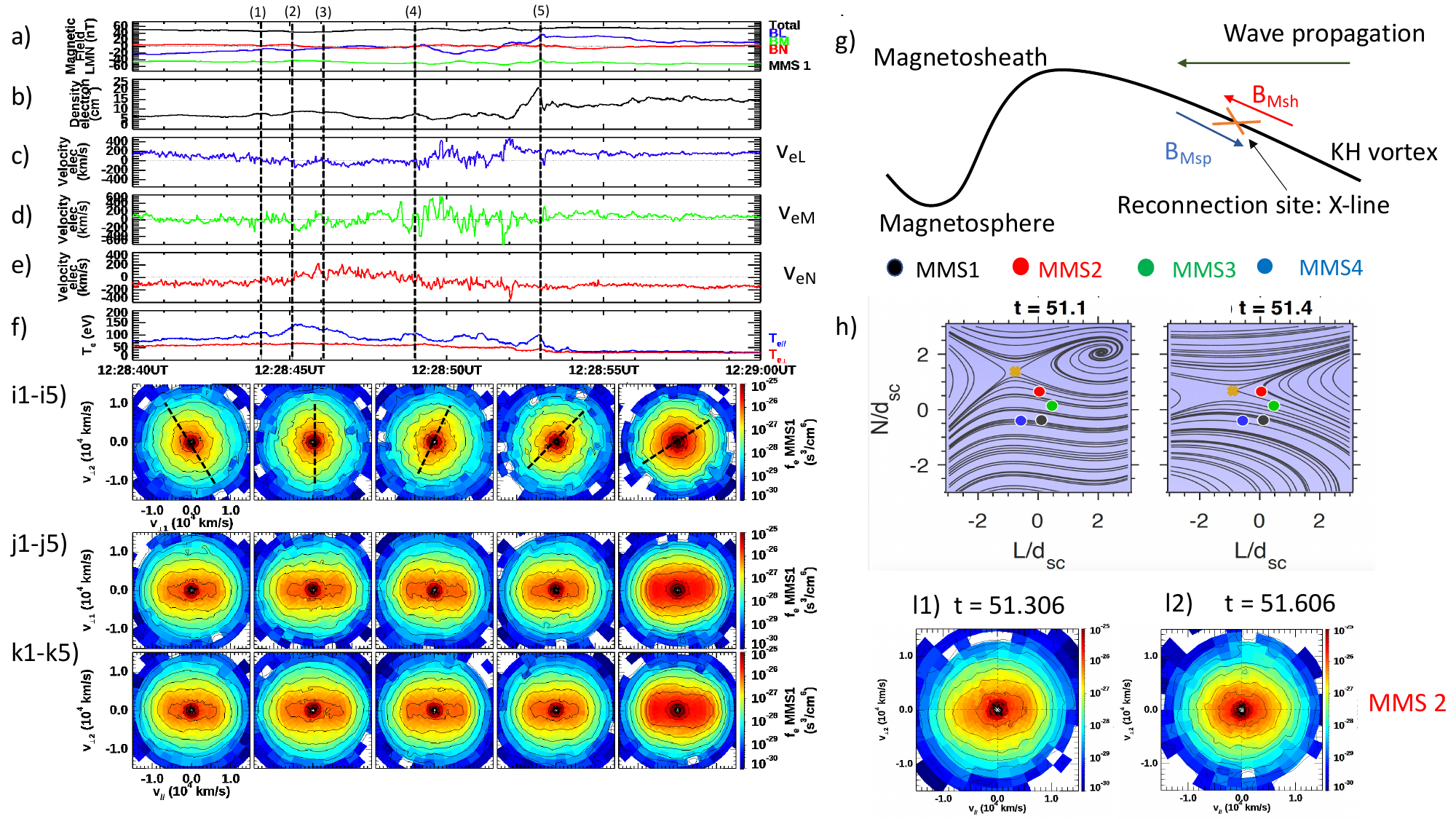}
\caption{MMS1 burst observations in local LMN at 12:28:40-12:29:00 UT on 14 April 2022. The panels show a) magnetic field components, b) electron density c), d), and e) electron velocity components $v_L$, $v_M$ and $v_N$, f) parallel and perpendicular temperatures of electrons. The bottom three rows of panels show the electron distribution functions at 5 different times corresponding to the dashed lines panels in (a-e) in the i$1$-i$5$) $v_{\bot1}- v_{\bot2}$ plane, j$1$-j$5$) the $v_{||}- v_{\bot1}$ plane and k$1$-k$5$) $v_{||}- v_{\bot2}$ plane. The electron distributions exhibit the anisotropy in the parallel direction, hence we see a temperature anisotropy of about $T_{e||}/T_{e\bot} \sim 3$. The eVDF at 12:28:45.100 exhibits agyrotropy, wherein the eVDF has a significant elongation in the $v_{\bot2}$ direction. g) Cartoon showing the reconnection site on the KH vortex, h) The reconstructed reconnection geometry using the method of \cite{Richard_Denton_2022} showing all the four spacecrafts. Between $t=51.1$ and $t=51.4$ the MMS2 spacecraft crosses the separatrix starting from the inflow to the exhaust. This crossing is evident in the eVDF at $t=51.306$ wherein we can observe an asymmetry in $v_{||}$, which indicates the mixing of magnetospheric and magnetosheath electron populations typically observed at the separatrix. Here $d_{sc}$ is the spacecraft separation. l1-l2) MMS2 eVDF at $t=51.306$ and $51.606$ in $v_{||}- v_{\bot2}$ plane.  
}
\label{Fig2:Stormtime_reconnectiondist}
\end{figure*}
Fig.~\ref{Fig2:Stormtime_reconnectiondist}a-f) shows 20 seconds of burst data of the magnetic field and the electron velocity field components in the local LMN frame. The boundary normal to the current sheet was estimated by applying minimum variance analysis on the magnetic field (MVA-B) observed by MMS1. 
The guide-field $\mathbf{M}$ direction is the unit vector along of the cross product between $\mathbf{M}$ and the maximum variance direction of $\mathbf{B}$ and $\mathbf{L} = \mathbf{M} \times \mathbf{N}$ completes the system. We performed MVA-B for interval: 12:28:49 to 12:28:53 yielding the directions of LMN as: $\mathbf{L}_{\scr{GSE}} = [-0.237, -0.823, 0.516]$, $\mathbf{M}_{\scr{GSE}} = [-0.215, -0.474, -0.854]$ and $\mathbf{N}_{\scr{GSE}} = [0.947, -0.313, -0.0646]$. During this 20s interval,  MMS-1 spans the boundary layer ($n_1=6cm^{-3}$) and magnetosheath ($n_2=20cm^{-3}$) with an average $d_i=c/\omega_{pi}=73$km, where c is the speed of the light and ion plasma frequency $\omega_{pi}=(n_1e^2/m_p\epsilon_0)$. The magnetic field showed a rotation from $B_{L1}=-20$nT to $B_{L2}=10$nT. This rotation was accompanied by a very weak (almost zero) $\mathbf{B}_N$ and a strong guide field $\mathbf{B}_M$ component of $50$nT. This corresponds to a guide field of $B_M/B_L=2.2$, supported by in-plane Hall electron currents. The current sheet associated with this rotation of magnetic field lasts for 5 seconds. During this interval, the spacecraft recorded an electron jet $v_{eL1}=200$km/s and a ion jet of the order $v_{iL}=100$km/s (not shown). The out-of-plane electron jets $v_{eM}$ experienced strong reversals along the current sheet of about $|v_{eM}|=500$km/s.

The local LMN coordinate system reported above was utilized to determine the reconnection geometry through polynomial reconstruction techniques developed by \cite{Richard_Denton_2022}. Fig.~\ref{Fig2:Stormtime_reconnectiondist}h) show the reconstructed reconnection geometry at time 12:28:51.1 and 12:28:51.4 UT respectively. From the reconstruction results, it can be inferred that MMS2 encounters the separatrix around 12:28:51.1 UT and crosses into the exhaust at 12:28:51.4 UT. Both MMS2 and MMS3 were observed to encounter the separatrix with a delay of $\Delta t=300 ms$. The separatrix crossing by MMS2 is also clearly evident in the electron distributions obtained by the DES instrument. At $51.3$, the electron distribution shows an asymmetry in $v_{||}$, where, $||$ refers to the direction parallel to the magnetic field. This asymmetry in $v_{||}$ indicates the mixing of two different electron populations, i.e in flowing magnetosheath electrons with the $v_{||} >0$ and exhaust electron populations with $v_{||}< 0$. All four MMS spacecraft start in the inflow region on the magnetosheath side, and MMS2 crosses the separatrix at  $51.306$. This is further evident by the strong temperature anisotropy observed in the electrons. Fig.~\ref{Fig2:Stormtime_reconnectiondist}g) shows the cartoon where the reconnection X-line is on the KH vortex, this corresponds to Type-I vortex induced reconnection(VIR). In Type-I VIR, the in-plane magnetic field is initially anti parallel across the velocity shear layer, and results in plasma mixing across the velocity shear layer along the reconnected field lines (\cite{li_kinetic_2016}). This reconnection occurs when the magnetopause current layer is locally compressed by the vortex flow resulting in the direct mixing of plasma across the shear layer.

The bottom three panels in Fig.~\ref{Fig2:Stormtime_reconnectiondist} show the electron distribution functions (eVDFs) at five different times in i1-i5) $v_{\bot1}- v_{\bot2}$ plane, j1-j5) $v_{||}- v_{\bot1}$ plane and k1-k5) $v_{||}- v_{\bot2}$ plane for MMS1,  where $\bot 1$ refers to the $(\mathbf{b} \times \mathbf{v})\times \mathbf{b}$ direction and $\bot 2$ refers to the $(\mathbf{b} \times \mathbf{v})$ direction. These distributions correspond to times 12:28:44.100, 12:28:45.100, 12:28:46.100, 12:28:49, and 12:28:53 UT from left to right marked as (1)-(5) in Fig.~\ref{Fig2:Stormtime_reconnectiondist} respectively. The eVDFs show a strong anisotropy in the parallel direction as shown in the bottom two rows of Fig.~\ref{Fig2:Stormtime_reconnectiondist} at all the marked times which is in agreement with the reconstruction results that suggest the MMS1 is in the inflow region of the reconnecting current sheet. The anisotropy in the electrons $T_{e||}/T_{e\bot}$ is about 3. Additionally, the eVDFs exhibit agyrotropy in the $v_{\bot1}$-$v_{\bot2}$ plane, in the form of elongation along $v_{\bot2}$ direction as demonstrated by the eVDF at 12:28:46.100 UT. The eVDFs at 12:28:44.100 and 12:28:49 UT also exhibit agyrotropy, however unlike the eVDF at  46.100 these eVDFs have elongation not purely along $v_{\bot2}$, but at an axis in $v_{\bot1}-v_{\bot2}$ plane. The three eVDFs exhibit a clockwise rotation in the agyrotropy. At times 12:28:49 and 12:28:53 UT the eVDFs become gyrotropic. During the 2 second interval 12:28:44.100 to 12:28:46.100 UT, when the electrons are agyrotropic, an out-of-plane electron jet $v_M=-200\mathrm{km/s} $ is observed. This jet is accompanied by a reversal in the normal component of the electron velocity $v_N$ around 12:28:49 (marked (4) in Fig.~\ref{Fig2:Stormtime_reconnectiondist}). Furthermore, using a $B_y$-minimum feature we estimated the structure velocity yielding $V_{\scr{nor}}=203\,\mathrm{km/s}$. During this encounter, the spacecraft separation is $d_{sc}=31km$. It was observed that the density ramp took place in about $\sim 1s$, giving us a length scale for the density gradient of about $203km$. 

Agyrotropic electrons have been previously observed by the MMS inside the electron diffusion region of an asymmetric reconnection at the day-side magnetopause(\cite{Egedal2016,Rager2018,Shuster2019}). These agyrotropic signatures result from strong gradients which often exist around the separatrices in asymmetric reconnection.  However those VDFs had a crescent-shaped distribution in the $v_{\bot1}$ - $v_{\bot2}$ plane, which arises due to spatial gradients in density and pressure. In contrast, the electron velocity distribution functions observed near the separatrix of reconnection at KH edges exhibit an "elliptical" agyrotropic feature, a key signature that has been not previously reported in KH studies. In the following section, we will quantify and discuss the physical implications of these agyrotropic signatures.\\

\section{Signatures of electron agyrotropy in Kelvin Helmholtz waves}\label{sec:Agyrotropy}
In regions where plasma properties like the density and the electron bulk velocity vary sharply, we observe strong agyrotropic distributions. Previous work has noted the relationship between agyrotropy and gradients at the scale of the Larmor radius (\cite{Scudder_2012,Scudder_2015}) primarily in reconnection scenarios, simple model equilibria with density gradients that lead to the gradients in the magnetized distribution function. In this section, we intend to analyze the observed agyrotropic distributions qualitatively and quantitatively. To quantify the agyrotropy in the electron VDFs we use the following scalar parameter introduced by (\cite{scudder_illuminating_2008}):

\begin{equation}
    A= 2\frac{|P_{\bot e1} - P_{\bot e2}|}{P_{\bot e1} + P_{\bot e2}}
    \label{Eqn1}
\end{equation}

\begin{figure*}
\centering
\includegraphics[width=\linewidth]{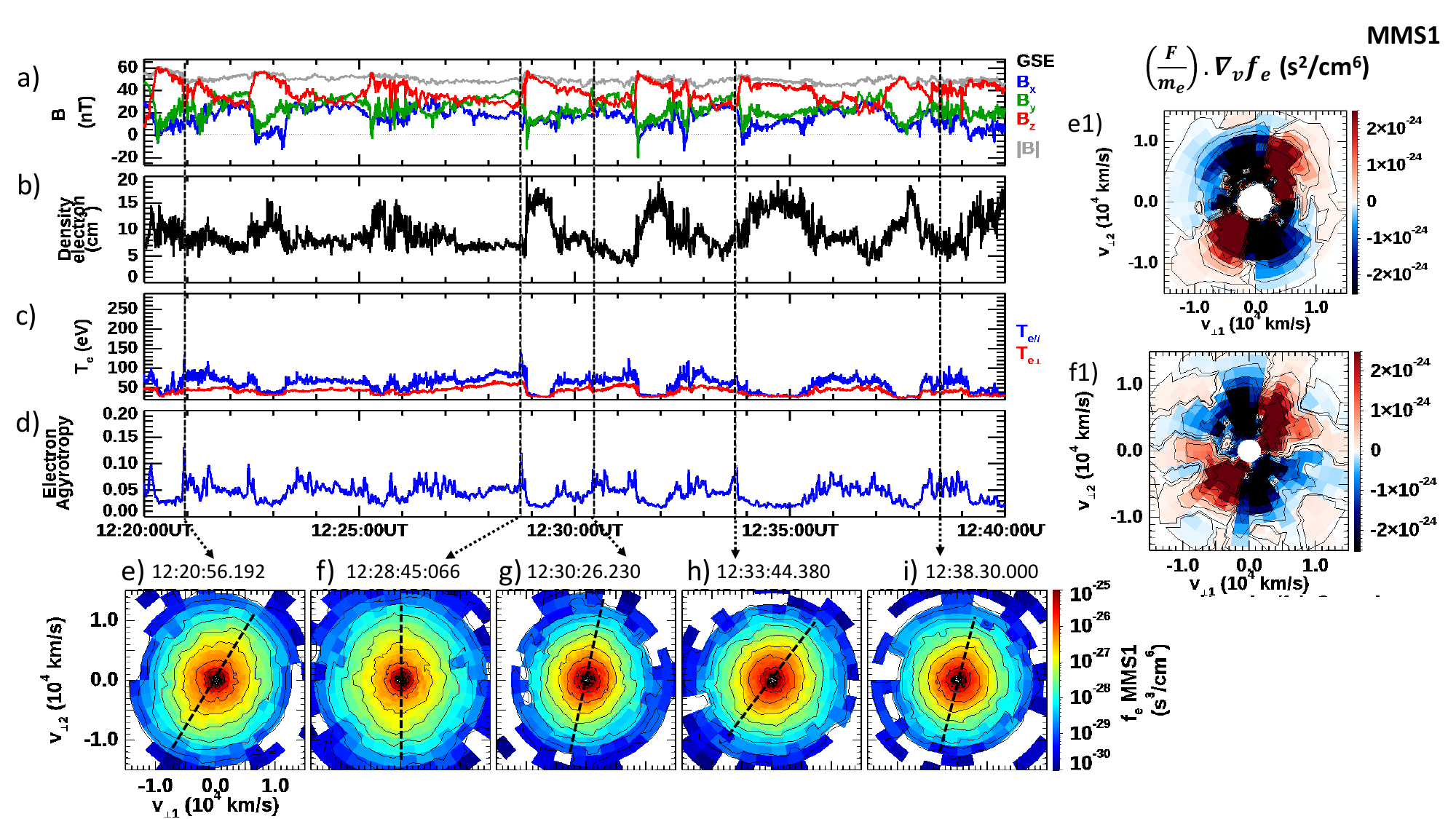}
\caption{ eVDFs exhibiting agyrotropy at the edges of KH vortices.  a) Magnetic field, b) density, c) parallel and perpendicular electron temperatures, and d) electron agyrotropy. The dashed lines mark the times with enhanced agyrotropy $A\sim 0.1$. e-i) eVDF's in the $v_{\bot1}- v_{\bot2}$ plane at these marked times. We used the definition provided by \cite{scudder_illuminating_2008} to measure the electron agyrotropy as given by Eqn.~\ref{Eqn1}. e1) and f1) MMS1 measurements of ${\bf v}\cdot\nabla f_e$ at 12:20:56.192 and 12:28:46.066 UT. The absence of electron counts near $|v|\leq 2000$kms$^{-1}$ in these MMS distributions is due to the subtraction of low-energy photo-electron contamination (\cite{Gershman_photoelec}).}
\label{Fig3:Stormtime_Nongyro_dist}
\end{figure*}
where $P_{\bot 1}$ and $P_{\bot 2}$ are pressures along $\bot 1$ and $\bot 2$ direction respectively.  This agyrotropy parameter measures the deviations of the VDFs in perpendicular velocity space from a circular shape, which is characteristic of gyrotropic plasmas. A value of $A=0$ indicates gyrotropic plasma, while an extreme value of 2 signifies highly non-gyrotropic plasma. Fig.~\ref{Fig3:Stormtime_Nongyro_dist} d) shows plot of the agyrotropy parameter during the KHI event. The plot shows spikes at the boundaries of KH waves, with an average agyrotropy measure of approximately 0.1 at these edges, in contrast to an average value of $0.015$ inside the vortices. This enhancement in agyrotropy by a factor of $10$ at the edges in the electron distributions is evident in Fig.~\ref {Fig3:Stormtime_Nongyro_dist} e-i) eVDFs which corresponds to times: 12:20:56.192, 12:28:45.066, 12:30:26.230, 12:33:44.380 and 12:38:30.000UT respectively.

In order to comprehend the origin of agyrotropy observed at the boundaries of Kelvin-Helmholtz waves, our analysis will primarily focus on examining the gradients in plasma parameters. It is important to note that the elliptical nature of the distribution indicates that diamagnetic effects are unlikely to be the driving force behind the observed electron agyrotropy. This is because the kinetic signature of diamagnetic effects typically manifests as crescent-shaped electron distributions, as demonstrated in previous studies (\cite{Egedal2016,Rager2018,Shuster2019}). To gain a deeper understanding of the kinetic origins of the elliptical agyrotropy, we refer to recent observations from the Magnetospheric Multiscale (MMS) mission and models that describe structures in terms of the electron Vlasov equation (\cite{shuster_structures_2021,shuster2023}). These works establish a fundamental connection between different types of bulk plasma gradients and their kinetic counterparts in the Vlasov equation. Specifically, they demonstrate that quadrupolar velocity-space signatures in the spatial gradient term ${\bf v}\cdot\nabla f_e$ are associated with spatial gradients in the bulk velocity $\nabla U_e$ (\cite{shuster_structures_2021}). For the specific KH event under investigation in this study, electron moments data from the DES is available for only three of the MMS spacecraft. As a result, it is not possible to directly measure ${\bf v}\cdot\nabla f_e$. Nonetheless, following the framework outlined in (\cite{shuster2023}), we can utilize single-spacecraft measurements of the velocity-space gradient term $({\bf F}/m_e)\cdot\nabla_{\bf v}f_e$ as a proxy for the spatial gradient term for quasi-steady-state structures in kinetic equilibrium moving rapidly past the spacecraft.

Fig.~\ref{Fig3:Stormtime_Nongyro_dist} e1) and f1) show characteristic examples of the Vlasov distribution structures  $({\bf F}/m_e)\cdot\nabla_{\bf v}f_e$ at time 12:20:56.192 and 12:28:46.066 UT close to the times Fig.~\ref{Fig3:Stormtime_Nongyro_dist} e) and f) respectively, capturing the occurrences of agyrotropy spikes. These 2D velocity-space slices show a pronounced quadrupolar pattern in $({\bf F}/m_e)\cdot\nabla_{\bf v}f_e$ that corresponds to and follows the rotating `propeller'-shaped elliptical structure of the electron distributions. Decomposing the $({\bf F}/m_e)\cdot\nabla_{\bf v}f_e$ signature into the contribution from the electric field $(-e/m_e){\bf E}\cdot\nabla_{\bf v}f_e$ and magnetic field $(-e/m_e)({\bf v}\times{\bf B})\cdot\nabla_{\bf v}f_e$ terms (not shown here), we can interpret the quadrupolar pattern by recognizing the dominant role played by the magnetic field term due to the substantial magnitude (a factor of $\sim3$) of $|{\bf B}|$. The term $(-{\bf v}\times{\bf B})\cdot\nabla_{\bf v}f_e$ is equivalent to the gyrophase derivative term: $B (\partial f_e / \partial \phi)$ (\cite{shuster2023}). Consequently, the elliptical agyrotropy observed in $f_e$ can be understood as an outcome of this underlying kinetic structure: going counter-clockwise around a circle of constant $|{\bf v}|$ in the $v_{\perp1}$-$v_{\perp2}$ plane, red regions corresponding to $(\partial f_e / \partial \phi) > 0$ result in an increase in phase space density, while blue regions with $(\partial f_e / \partial \phi) < 0$ indicate a decrease in phase space density. The presence of quadrupolar signatures in $({\bf F}/m_e)\cdot\nabla_{\bf v}f_e$ imply the existence of electron-spatial-scale gradients in the electron bulk velocity, consistent with previous findings. This is a reasonable inference since the overall KH structure arises from larger ion-scale vortices, making it plausible for smaller-scale embedded bulk velocity shears to develop at these electron scales, which can now be probed by MMS instrumentation.

\section{Summary and Discussions} \label{sec:Final}

In this paper, we explore the key physics underlying turbulence and reconnection within KH vortices during a geomagnetic storm, utilizing observations measured by MMS. Our analysis shows the presence of power-law spectra in both electric and magnetic field data, indicating the occurrence of turbulence. Notably, while the magnetic field power spectra demonstrate expected changes in spectral slope near the ion gyrofrequency, the electric field spectra do not exhibit a break at the ion cyclotron frequency; instead, a break is observed near the lower hybrid frequency. Furthermore, we observe shallower spectral slopes compared to turbulence in magnetotail region, suggesting a slower dissipation to small scales, possibly attributable to the storm conditions. The storm conditions may lead to increased energy buildup, delaying dissipation to smaller scales. We are currently engaged in further research to comprehensively understand turbulence generation within KH waves under both storm and non-storm conditions.


During this event, the MMS spacecraft observed a current sheet located approximately $18d_e$($\sim 0.6d_{sc}$) away from an electron diffusion region, accompanied by intense electron jets. This aligns with features associated with strong guide-field asymmetric reconnection occurring across the magnetopause. Alongside a strong anisotropy ($T_{||}>T_{\bot}$) in the inflow region the electrons show deviations from gyrotropic behavior near the reconnecting current sheet. This agyrotropy was observed at the edges of the KH vortices, where pronounced magnetic gradients are prominent. The measured agyrotropy was approximately $0.1$. This agyrotropy manifests as an elliptical structure in electron velocity distribution functions, distinct from the crescent distributions extensively observed in MMS data. While the crescent shaped distributions arise from density gradients, this agyrotropy arises from the gradients in electron bulk flow velocity that arise possibly due to the intense storm-time conditions or the large velocity shears at electron scales inside the vortices. Further examination of other KH encounters by MMS revealed similar agyrotropy in two additional events, one occurring during a storm and the other under non-storm conditions. Consequently, the source of this elliptical agyrotropy remains unanswered within the realm of KH studies, prompting further investigation to determine the magnitude of electron-scale gradients required for agyrotropy manifestation within reconnecting current sheets, while also exploring the influence of these reconnecting current sheets on KH wave power spectra.\\

We extend our appreciation to the MMS team for providing the data used in this study, sourced from the MMS Science Data Center \url{(https://lasp.colorado.edu/mms/sdc/)}. MRA was supported by NASA grant 80NSSC18K1359.

\bibliography{main_ref}{}
\bibliographystyle{aasjournal}

\end{document}